# Fair Isaac Technical Paper

**Subject:** **Score Engineered Logistic Regression**

**From:** **Bruce Hoadley (BCH – A34 X27051)**

**Date:** **September 13, 2000**

## Abstract

In several Fair Isaac studies logistic regression has been shown to be a very competitive technology for developing unrestricted scoring models – especially for performance metrics like ROC area. Application of logistic regression has been hampered by the lack of software to handle complex score engineering, such as pattern constraints. The purpose of this paper is to develop a sequential quadratic programming approach to score engineered logistic regression. Gerald Fahner, Reference [6], has developed in SAS an approach to score engineered logistic regression based on iterative re-weighted least squares. This is the method used in SAS proc logistic. Gerald just overlaid constrained least squares to handle the score engineering. The sequential quadratic programming approach is based on a simple Taylor series expansion of minus log likelihood, which is locally quadratic. This is a more direct method for solving the problem. And it fits in with the methods described in References [1], [3], and [5], which are all based on quadratic programming or sequential quadratic programming. The paper also provides all of the simple MATLAB code for implementing the algorithm. In the same large score engineered fraud test example used in References [1], [3], and [5], the algorithm converged after four iterations. And, each iteration took 12 seconds.

There are several problems where logistic regression might be useful to Fair, Isaac. One problem is the scorecard with only a few characteristics and a small handful of attributes. In this case, the score distributions may be non-Normal, so that the divergence objective function is iffy. A second case is where the primary objective is to develop a very accurate estimate of the probability of Good for each account, rather than a score for making binary decisions. A third application is Object*Boost*, where the objective is to find observation weights so that the logistic regression score optimizes some business objective on the validation sample. To get a score engineered version of Object*Boost*, we need score engineered logistic regression. The Object*Boost* concept also works with scores that maximize divergence, but not as well as logistic regression. This is because divergence maximization is not as sensitive to observation weights.

# Table of Contents

## 1. Classical Logistic Regression

In logistic regression we have a binary outcome variable, $y$, which can have the values 0 and 1, and a set of numerical predictors, $\boldsymbol{x}' = (x_1, \ldots, x_p)$. By tradition, $x_1 \equiv 1$. This accommodates an intercept term in the model.

Let

$$p(\boldsymbol{x}) = \Pr\{y = 1 \mid \boldsymbol{x}\}.$$

There are two key assumptions underling logistic regression.

1. $y \mid \boldsymbol{x} \sim \text{Bernoulli}[p(\boldsymbol{x})]$
2. $\log\left(\dfrac{p(\boldsymbol{x})}{1 - p(\boldsymbol{x})}\right) = \sum_{j=1}^{q} \beta_j x_j$ .

The $\beta_j$'s are called regression coefficients. The coefficient, $\beta_1$, is an intercept term. For scorecards and liquid scorecards (see Reference [3]) the $x_j$'s for $j = 2, \ldots, q$ are either attribute indicator variables or B-spline basis functions. And since the liquid scorecard is so general, this pretty much covers the waterfront.

Our development data consists of $n$ weighted observations on the variables $(y, \boldsymbol{x})$:

$$(y_1, \boldsymbol{x}_1), \ldots, (y_n, \boldsymbol{x}_n).$$

The weight is

$$\boldsymbol{w}' = (w_1, \ldots, w_n) .$$

The Bernoulli likelihood function is

$$\prod_{i=1}^{n} [p(\boldsymbol{x}_i)]^{w_i y_i} [1 - p(\boldsymbol{x}_i)]^{w_i (1 - y_i)}.$$

The log likelihood is

$$\begin{aligned} L &= \sum_{i=1}^{n} \{w_i y_i \log[p(\boldsymbol{x}_i)] + w_i (1 - y_i) \log[1 - p(\boldsymbol{x}_i)]\} \\ &= \sum_{i=1}^{n} w_i \left\{ y_i \log\left[\frac{p(\boldsymbol{x}_i)}{1 - p(\boldsymbol{x}_i)}\right] + \log[1 - p(\boldsymbol{x}_i)] \right\}. \end{aligned}$$

It is useful to define the parameter

$$\theta_i = \log\left(\frac{p(\boldsymbol{x}_i)}{1 - p(\boldsymbol{x}_i)}\right) = \sum_{j=1}^{q} \beta_j x_{ij} \; .$$

A little algebra reveals the relationships

$$p(\boldsymbol{x}_i) = \frac{\exp(\theta_i)}{1 + \exp(\theta_i)}$$
$$1 - p(\boldsymbol{x}_i) = \frac{1}{1 + \exp(\theta_i)}$$
$$\log[1 - p(\boldsymbol{x}_i)] = -\log[1 + \exp(\theta_i)].$$

Minus log likelihood can be written as

$$\begin{aligned} M(\beta) &= -L(\beta) \\ &= \sum_{i=1}^{n} - w_i \{y_i \theta_i - \log[1 + \exp(\theta_i)]\}. \end{aligned}$$

This minus log likelihood depends on the regression coefficients, $\beta' = (\beta_1, \ldots, \beta_q)$, because

$$\theta = \boldsymbol{X} * \beta,$$

where

$$\theta' = (\theta_1, \ldots, \theta_n)$$
$\boldsymbol{X} = n \times q$ design matrix with elements, $x_{ij}$ .

The classical logistic regression problem is

Find $\beta$ to
Minimize $M(\beta)$.

This is the problem solved by proc logistic in SAS, which uses iterative re-weighted least squares (see Chapter 13 of Reference [4]).

## 2. Score Engineered Logistic Regression

Classical logistic regression has limited applicability for Fair, Isaac and Company Inc., because it does not handle score engineering. To describe the score engineering problem it is convenient to decompose $\beta$ into

$$\beta' = [S_0 \quad \boldsymbol{S}'],$$

where $S_0$ is the intercept coefficient and $\boldsymbol{S}$ are the score weights associated with the scorecard or liquid scorecard.

With score engineering, the logistic regression problem is

$$\begin{aligned}&\text{Find } \beta \text{ to}\\ &\text{Minimize } \; M(\beta) + \frac{\lambda}{q-1}\boldsymbol{S}' * \boldsymbol{S}\\ &\text{Subject to:}\\ &\quad \boldsymbol{Aeq} * \beta = \boldsymbol{beq}\\ &\quad \boldsymbol{A} * \beta \le \boldsymbol{b}.\end{aligned}$$

The first set of constraints is the centering, cross restriction, group restriction, and in-weighting constraints. Most of these constraints involve only the scorecard part of the model, but you might want to in-weight $S_0$ to, e.g., the empirical log pop odds. The second set of constraints is the pattern constraints, which typically only involve the scorecard part of the model.

Note that in the logistic regression case, there is no weight of evidence scale. This is because the dependent variable, $y$, dictates the scale of the final model. However, the scorecard part of the model is a rough model of the information odds, which is approximately on the weight of evidence scale. And the score weights for the scorecard part of the model will vary around zero. This is why the penalty term involves only the scorecard part of the model.

The score engineered logistic regression problem can be re-expressed by defining the matrix

$$\boldsymbol{Ir} = \begin{bmatrix} 0 & 0 & 0 & 0 \\ 0 & 1 & 0 & 0 \\ 0 & 0 & 1 & 0 \\ 0 & 0 & 0 & 1 \end{bmatrix} \quad (q \times q \text{ matrix}) .$$

In this notation, the score engineered logistic regression problem is

$$\begin{aligned} &\text{Find } \beta \text{ to} \\ &\text{Minimize} \quad M(\beta) + \frac{\lambda}{q-1} \beta' * \boldsymbol{Ir} * \beta \\ &\text{Subject to:} \\ &\quad \boldsymbol{Aeq} * \beta = \boldsymbol{beq} \\ &\quad \boldsymbol{A} * \beta \le \boldsymbol{b} . \end{aligned}$$

This is starting to look like the quadratic programming problems described in Reference [1]. However, $M(\beta)$ is not a quadratic function. So the score engineered logistic regression problem is a non-linear programming problem.

## 3. A Quadratic Programming Formulation

The classical logistic regression problem is solved by iterative re-weighted least squares; i.e., a sequence of quadratic problems, Reference [4]. This method can be adapted to the score engineered case, by formulating an appropriate sequence of quadratic programs. In Reference [1] I showed that score development problems could be solved very quickly via quadratic programming.

Let $\hat{\beta}$ denote the current best solution to the problem. Via Taylor series, I can approximate minus log likelihood in the neighborhood of $\hat{\beta}$ by the quadratic function

$$\hat{M}(\beta) = M(\hat{\beta}) + \boldsymbol{g}' * (\beta - \hat{\beta}) + \frac{1}{2} (\beta - \hat{\beta})' * \boldsymbol{G} * (\beta - \hat{\beta}),$$

where $\boldsymbol{g}$ and $\boldsymbol{G}$ are the gradient vector and Hessian matrix of the function, $M(\beta)$, evaluated at $\beta = \hat{\beta}$.

With this approximation the problem becomes

Find $\beta$ to

Minimize $\boldsymbol{g}' * (\beta - \hat{\beta}) + \frac{1}{2}(\beta - \hat{\beta})' * \boldsymbol{G} * (\beta - \hat{\beta}) + \frac{\lambda}{q-1}\beta' * \boldsymbol{Ir} * \beta$

Subject to:

$\boldsymbol{Ai} * \beta = \boldsymbol{IW}$

$\boldsymbol{Acr} * \beta = 0$

$\boldsymbol{Apr} * \beta \le 0$.

The objective function can be manipulated as

$$
\begin{aligned}
&\boldsymbol{g}' * (\beta - \hat{\beta}) + \frac{1}{2}(\beta - \hat{\beta})' * \boldsymbol{G} * (\beta - \hat{\beta}) + \frac{\lambda}{q-1}\beta' * \boldsymbol{Ir} * \beta \\
&= \boldsymbol{g}' * \beta - \boldsymbol{g}' * \hat{\beta} + \frac{1}{2}\left[\beta' * \boldsymbol{G} * \beta - 2\hat{\beta}' * \boldsymbol{G} * \beta + \hat{\beta}' * \boldsymbol{G} * \hat{\beta}\right] \\
&\quad + \frac{\lambda}{q-1}\beta' * \boldsymbol{Ir} * \beta \\
&= \frac{1}{2}\hat{\beta}' * \boldsymbol{G} * \hat{\beta} - \boldsymbol{g}' * \hat{\beta} + \left[\boldsymbol{g}' - \hat{\beta}' * \boldsymbol{G}\right] * \beta + \beta' * \frac{1}{2}\left[\boldsymbol{G} + \frac{2\lambda}{q-1}\boldsymbol{Ir}\right] * \beta .
\end{aligned}
$$

Now the problem becomes

Find $\beta$ to

Minimize $\frac{1}{2}\beta' * \left[\boldsymbol{G} + \frac{2\lambda}{q-1}\boldsymbol{Ir}\right] * \beta + \left[\boldsymbol{g}' - \hat{\beta}' * \boldsymbol{G}\right] * \beta$

Subject to:

$\boldsymbol{Aeq} * \beta = \boldsymbol{beq}$

$\boldsymbol{A} * \beta \le \boldsymbol{b}$.

**Derivation of $g$ and $G$**

By definition

$$
\boldsymbol{g}' = \left(\frac{\partial M(\beta)}{\partial \beta_1}, \ldots, \frac{\partial M(\beta)}{\partial \beta_q}\right)
$$

$$
\boldsymbol{G} = \left[\frac{\partial^2 M(\beta)}{\partial \beta_h \partial \beta_k}\right] \quad \text{a } q \times q \text{ symmetric matrix},
$$

where these are evaluated at $\beta = \hat{\beta}$.

Let

$$m_i = -w_i \{ y_i \theta_i - \log[1 + \exp(\theta_i)] \}$$

so that I can write minus log likelihood as

$$M(\beta) = \sum_{i=1}^{n} m_i \ .$$

This means that

$$\frac{\partial M(\beta)}{\partial \beta_k} = \sum_{i=1}^{n} \frac{\partial m_i}{\partial \beta_k}$$
$$\frac{\partial^2 M(\beta)}{\partial \beta_h \partial \beta_k} = \sum_{i=1}^{n} \frac{\partial^2 m_i}{\partial \beta_h \partial \beta_k} \ .$$

The chain rule and elementary calculus yield

$$\begin{aligned} \frac{\partial m_i}{\partial \beta_k} &= \frac{\partial m_i}{\partial \theta_i} \frac{\partial \theta_i}{\partial \beta_k} \\ &= \left[ -w_i \left\{ y_i - \frac{\exp(\theta_i)}{1 + \exp(\theta_i)} \right\} \right] [x_{ik}] \\ &= [-w_i \{ y_i - p_i \}] \bullet [x_{ik}] \\ &= w_i (p_i - y_i) x_{ik} \end{aligned}$$

Applying the chain rule again yields

$$\frac{\partial^2 m_i}{\partial\beta_h \partial\beta_k} = \frac{\partial\left[\dfrac{\partial m_i}{\partial\beta_k}\right]}{\partial\theta_i}\frac{\partial\theta_i}{\partial\beta_h}$$
$$= \left[w_i\left(\frac{\exp(\theta_i)}{1+\exp(\theta_i)}\right)\left(\frac{1}{1+\exp(\theta_i)}\right)\right]\bullet[x_{ik}]\bullet[x_{ih}]$$
$$= w_i p_i(1-p_i)x_{ih}x_{ik}\ .$$

Hence

$$\frac{\partial M(\beta)}{\partial\beta_k} = \sum_{i=1}^{n} w_i(p_i - y_i)x_{ik}$$
$$\frac{\partial^2 M(\beta)}{\partial\beta_h\partial\beta_k} = \sum_{i=1}^{n} w_i p_i(1-p_i)x_{ih}x_{ik}\ .$$

**Matrix Computation of $\boldsymbol{g}$, $\boldsymbol{G}$, and $\boldsymbol{M}$**

To write the gradient and Hessian in matrix notation define

$$\boldsymbol{p}' = (p_1,\ldots,p_n)$$
$$\boldsymbol{y}' = (y_1,\ldots,y_n)$$
$$\boldsymbol{w}' = (w_1,\ldots,w_n)$$
$\boldsymbol{Wp} =$ $n \times n$ diagonal matrix with $Wp_{ii} = w_i p_i(1-p_i)$.

In Section 1, I noted that

$$\theta = \boldsymbol{X} * \beta$$
$$\boldsymbol{p} = \exp(\theta)\bullet/(\boldsymbol{1}+\exp(\theta)),$$

where $\bullet/$ means element by element division.

Then

$$\boldsymbol{g} = \boldsymbol{X}' * [\boldsymbol{w}\bullet*(\boldsymbol{p}-\boldsymbol{y})]$$
$$\boldsymbol{G} = \boldsymbol{X}' * \boldsymbol{Wp} * \boldsymbol{X}$$
$$M = \boldsymbol{w}' * [\log(1+\exp(\theta)) - \boldsymbol{y}\bullet*\theta],$$

where $*$ is matrix multiplication and $\bullet*$ is array element by element multiplication.

**MATLAB Formulation**

For the score engineered logistic regression quadratic program, the matrices in the general form of the MATLAB quadratic program (see Section 2.2 of Reference [1]) are

$$H = G + \frac{2\lambda}{q-1} Ir$$

$$f = g - G * \hat{\beta}$$

$$Aeq, beq, A, b$$

$$l = -\infty$$

$$u = +\infty .$$

# 4. Application

To test the methodology, I will use the fraud test case used in References [1]. In this example, there are 25 prediction characteristics and 171 scorecard attributes. So the score engineered logistic regression model has 172 score coefficients, because there is an intercept term.

**Iteration of the algorithm**

The function, **beta**, given in Appendix 1 performs one iteration of the algorithm for computing the score engineered logistic regression solution. The current value of the logistic regression coefficients is called betain. The new value of the logistic regression coefficients is called betaout. The function is based on the theory developed above.

The initial iteration of the algorithm is the MATLAB code

```
betaout=beta(Xr,yd,w,0,A,b,Aeq,beq,l,u,beta0);
```

I describe below how each of the input terms is computed.

**Design matrix**

From a design matrix point of view, logistic regression is the same as least squares regression, which was documented in Section 6 of Reference [1]. So I use the same design matrix as was used there. The computation of `Xr` is shown on p. 39 of Reference [1].

**Performance variable**

The performance variable is the same one that was used in several of the score developments in Reference [1]. The computation of yd is shown on p.16 of Reference [1].

**Observation weights**

For comparison purposes, I will use the same observation weights used in Reference [1]; i.e., equal weights. So the MATLAB code is

```
w=ones(9907,1);
```

**Penalty parameter**

For comparison purposes, I will use the same penalty parameter used for the development of S1 in Section 2 of Reference [1]; i.e. $\lambda = 0$. So the MATLAB code is

```
lambda=0;
```

**Constraint matrices for the quadratic program**

For this application I will use the 59 equality constraints used for S1 in Section 2.5 of Reference [1]. Hence, in MATLAB code

```
Aeq=[zeros(59,1) Ac] ;
```

where `Ac` was defined in Section 2.5 of Reference [1]. The first column of `Aeq` reflects the existence of the additional intercept score coefficient.

In this example, there is no non-zero in-weighting, hence in MATLAB code

```
beq=zeros(59,1);
```

For this application I will use the 106 pattern constraints used for S1 in Section 2.5 of Reference [1]. Hence, in MATLAB code

```
A=[zeros(106,1) Ap] ;
```

where `Ap` was defined in Section 2.5 of Reference [1]. The first column of `A` reflects the existence of the additional intercept score coefficient.

In this example, all of the inequality constraints are pattern constraints, hence in MATLAB code

```
b=zeros(106,1);
```

And as usual (in all my papers on quadratic programming),

```
l=-inf*ones(172,1);
u=inf*ones(172,1);
```

**Initial value of $\beta$**

As an initial value of $\beta$ I will use the scorecard, S1, in Reference [1] as the scorecard part of the logistic regression model. This is appropriated, because this scorecard is on a weight of evidence scale. For the initial intercept term, I will use log pop odds.

For the development sample there are 4940 Goods and 4967 Bads and $w_i \equiv 1$. So log pop odds is about equal to 0.

In MATLAB code the initial value of $\beta$ is

```
beta0=[0;S1];
```

**Running the algorithm manually**

A manual run of the algorithm was done as follows

```
betaout=beta(Xr,yd,w,0,A,b,Aeq,beq,l,u,beta0);
betaout2=beta(Xr,yd,w,0,A,b,Aeq,beq,l,u,betaout);
betaout3=beta(Xr,yd,w,0,A,b,Aeq,beq,l,u,betaout2);
betaout4=beta(Xr,yd,w,0,A,b,Aeq,beq,l,u,betaout3);
```

Each time I measured how close I was to convergence by

```
max(abs(betaout-betain))
```

Here is a table of the results

**Results of Logistic Regression Iteration**

| Iteration | betain | betaout | max(abs(betaout-betain)) |
|---|---|---|---|
| 1 | beta0 | betaout | .66 |
| 2 | betaout | betaout2 | .18 |
| 3 | betaout2 | betaout3 | .013 |
| 4 | betaout3 | betaout4 | .000057 |

Amazingly, it converged in four iterations.

**Performance results**

In this section I compare two engineered scores. The first engineered score is the score that maximizes divergence. The MATLAB computation of this score is

```
Score=Xd*S1;
```

where `Xd` and `S1` are defined in Section 2.5 of Reference [1]. This score is also given in Section 4.3 of Reference [5]. In fact, the matrices, `Xd` and `S1` are in the MATLAB dataset, marginal, associated with Reference [5].

The second engineered score is the logistic regression score, which was designed to minimize the minus log likelihood function. It is defined by the MATLAB code

```
score4=Xr*betaout4;
```

The table below shows how they fared

| Engineered Score | Development Divergence | Development - Log Likelihood |
|---|---|---|
| Score (max divergence) | 1.753 | 5070 |
| score4 (logistic regression) | 1.716 | 5046 |

The MATLAB code for computing the first column of this table is

```
Divergscore(Score,yd)
Divergscore(score4,yd)
```

The function, **Divergscore**, is documented in Appendix 2 of Reference [5].

The MATLAB code for computing the second column of this table is

```
MLRLL(Xr,yde,w,beta0)
MLRLL(Xr,yd,w,betaout4)
```

The function, **MLRLL**, is documented in Appendix 1.

These results are consistent with what you would expect. The engineered score, Score, has higher divergence and the engineered score, score4, has less minus log likelihood. And both scores satisfy all the score engineering constraints.

We can also compare these scores on measures based on the ROC curve, even though the ROC curve was not used in their developments.

| Engineered Score | Development KS | Development ROC Area |
|---|---|---|
| Score (max divergence) | .4964 | .8250 |
| score4 (logistic regression) | .4973 | .8257 |

The MATLAB code for computing the first row of this table is

```
[KS,ROCA]=ROC(Score,yd)
```

The function, **ROC**, is documented in Appendix 1. This function uses a new function, **scorecdfs**, which is also documented in Appendix 1.

The MATLAB code for computing the second row of this table is

```
[KS4,ROCA4]=ROC(score4,yd)
```

Here we see that the logistic regression score has slightly better ROC curve properties on the development sample. This is a typical result.

The development ROC area result for my engineered logistic regression score is virtually the same as that obtained by Gerald Fahner using iterative re-weighted least squares, Reference [6]. The ROC area result for Gerald's score is documented in Reference [7], where his score is called IRCLS (Iterative Re-weighted Constrained Least Squares).

These kinds of development sample results usually hold true for the validation sample for this kind of data and scorecard. However, I do not provide the validation sample results here, because that is not the purpose of this paper.

**Logistic Regression Coefficients**

The intercept coefficient is -.1026. The fact that this is not closer to the log pop odds of near zero, may reflect the constraints of score engineering.

The rest of the coefficients are shown in Appendix 2, column 6. For comparison, column 5 shows the score weights for the weight of evidence score, which maximizes divergence. As you can see, the coefficients are similar, but there are some differences. However, all of the score engineering constraints are satisfied for the logistic regression score weights.

Column 7 shows the score engineered logistic regression solution obtained by Gerald Fahner using the iterative re-weighted constrained least squares (IRCLS) approach to logistic regression, Reference [6]. As you can see, the solutions agree for the most part in the first two decimal places. Gerald's intercept term was -.0920, which is close to my value of -.1026.

## Appendix 1. New MATLAB Functions

This appendix documents additional MATLAB functions that I wrote to do this analysis. This adds to the collection of functions documented in Appendix 2 of Reference [3] and Appendix 2 of Reference [5].

**beta**

```
function ...
betaout=beta(X,y,w,lambda,A,b,Aeq,beq,l,u,betain)

% This function computes one iteration of the
%   logistic regression interative quadratic
%   programming algorithm.

%   X = design matrix
%   y = binary outcome vector
%   w = observation weight vector
%   lambda = penalty parameter
%   A,b,Aeq,beq,l,u are the constraint
%     matrices and vectors
%   betain is the input value of the logistic
%      regression coefficient vector
%   betaout is the output value of the logistic
%      regression coefficient vector
```

```
   q=length(betain);
   n=length(y);
   one=ones(n,1);
   Ir=eye(q);
   Ir(1,1)=0;
   theta=X*betain;
   p=exp(theta)./(one+exp(theta));
   g=X'*(w.*(p-y));
   wp=w.*p.*(one-p);
   Wp=spdiags(wp,0,n,n);
   G=X'*Wp*X;
   H=G+(2*lambda/(q-1))*Ir;
   f=g-G*betain;
   betaout=quadprog(H,f,A,b,Aeq,beq,l,u,betain);
```

**MLRLL**

```
function M=MLRLL(X,y,w,beta)

% This function computes minus
%   logistic regression log likelihood

%   X = design matrix
%   y = binary outcome vector
%   w = observation weight vector
%   beta =  logistic regression coefficient vector

   n=length(y);
   theta=X*beta;
   M=w'*(log(ones(n,1)+exp(theta))-y.*theta)
```

**scorecdfs**

```
function [FG,FB,orscore]=scorecdfs(score,y)

% This function computes the ordered score
%    and the cdfs of the Goods and Bads

% length of score and y has to be equal
% score is a vector of scores
% y is a vector of binary performances, 1 is Good
% FG is the vector: cdf of the Goods
% FB is the vector: cdf of the Bads
% orscore is the ordered score vector

n=length(y);
A=[score y];
B=sortrows(A,1);
orscore=B(:,1);
CG=cumsum(B(:,2));
CB=cumsum(ones(n,1)-B(:,2));
FG=CG/CG(n);
FB=CB/CB(n);
```

**ROC**

```
function [KS,ROCA]=ROC(score,y)

% This function computes the KS statistic
%   and the ROC area

% length of score and y has to be equal
% score is a vector of scores
% y is a vector of binary performances, 1 is Good
% KS is the Kolmogorov-Smirnoff statistic
% ROCA is the ROC area

[FG,FB,orscore]=scorecdfs(score,y);
KS=max(FB-FG);
ROCA=trapz(FG,FB);
```

# Appendix 2. Score Engineered Scorecards

| Char | Attribute | Att # | Constraint | Score Weights | | |
|---|---|---|---|---|---|---|
| | | | | Maximum Divergence | QP Logistic Regression | IRCLS Logistic Regression |
| char170 | -9999999 | 1 | " = 0 " | 0.000 | -0.0000 | 0.0000 |
| char170 | 0-<5 | 2 | > 3 | 0.306 | 0.3486 | 0.3465 |
| char170 | 5-<25 | 3 | > 4 | 0.157 | 0.1596 | 0.1591 |
| char170 | 25-<35 | 4 | > 5 | -0.067 | -0.0978 | -0.0972 |
| char170 | 35-<300 | 5 | > 6 | -0.259 | -0.2767 | -0.2754 |
| char170 | 300-High | 6 | | -0.888 | -0.8210 | -0.8192 |
| char170 | NO INFORMATION | 7 | " = 0 " | 0.000 | 0.0000 | 0.0000 |
| char191 | -9999999 | 8 | " = 0 " | 0.000 | -0.0000 | 0.0000 |
| char191 | 0-<2 | 9 | < 10 | -1.150 | -1.5565 | -1.5322 |
| char191 | 2-<5 | 10 | < 11 | -1.088 | -0.9659 | -0.9639 |
| char191 | 5-<7 | 11 | < 12 | -0.630 | -0.6835 | -0.6803 |
| char191 | 7-<650 | 12 | < 13 | 0.000 | 0.0037 | 0.0033 |
| char191 | 650-High | 13 | | 0.096 | 0.1047 | 0.1041 |
| char191 | NO INFORMATION | 14 | " = 0 " | 0.000 | -0.0000 | 0.0000 |
| char193 | -9999999 | 15 | | 0.396 | 0.4136 | 0.4120 |
| char193 | 0 | 16 | < 17 | -1.485 | -1.5459 | -1.5397 |
| char193 | 1 | 17 | < 18 | -1.317 | -1.3312 | -1.3275 |
| char193 | 2 | 18 | < 19 | -1.262 | -1.2254 | -1.2214 |
| char193 | 3-<18 | 19 | < 20 | -0.086 | -0.0898 | -0.0895 |
| char193 | 18-High | 20 | | 0.038 | 0.0383 | 0.0382 |
| char193 | NO INFORMATION | 21 | " = 0 " | 0.000 | -0.0000 | 0.0000 |
| char211 | -9999999 | 22 | | -0.096 | -0.0400 | -0.0403 |
| char211 | -9999998 | 23 | | 0.545 | 0.4579 | 0.4570 |
| char211 | 0 | 24 | < 31 | 0.449 | 0.4003 | 0.3998 |
| char211 | 1-<7 | 25 | < 30 | -0.064 | -0.0054 | -0.0068 |
| char211 | 7-<35 | 26 | < 27 | -0.916 | -0.8751 | -0.8722 |
| char211 | 35-<80 | 27 | < 28 | -0.674 | -0.6259 | -0.6236 |
| char211 | 80-<200 | 28 | < 29 | -0.392 | -0.3086 | -0.3083 |
| char211 | 200-<400 | 29 | < 30 | -0.064 | -0.0054 | -0.0068 |
| char211 | 400-<800 | 30 | < 31 | -0.064 | -0.0054 | -0.0068 |
| char211 | 800-<1300 | 31 | < 32 | 0.449 | 0.4003 | 0.3998 |
| char211 | 1300-<1700 | 32 | < 33 | 0.449 | 0.4003 | 0.3998 |
| char211 | 1700-High | 33 | | 0.449 | 0.4003 | 0.3998 |
| char211 | NO INFORMATION | 34 | " = 0 " | 0.000 | 0 | 0.0000 |
| char260 | -9999999 | 35 | | -0.162 | -0.1619 | -0.1613 |
| char260 | 0-<101 | 36 | > 37 | 0.463 | 0.4377 | 0.4367 |
| char260 | 101-<210 | 37 | > 38 | 0.314 | 0.3229 | 0.3214 |
| char260 | 210-<305 | 38 | > 39 | 0.312 | 0.3229 | 0.3214 |
| char260 | 305-<565 | 39 | > 40 | 0.143 | 0.1998 | 0.1976 |
| char260 | 565-<700 | 40 | > 41 | -0.276 | -0.3260 | -0.3232 |
| char260 | 700-High | 41 | | -0.276 | -0.3260 | -0.3232 |
| char260 | NO INFORMATION | 42 | " = 0 " | 0.000 | 0 | 0.0000 |
| char320 | -9999999-<0 | 43 | > 44 | 0.366 | 0.3619 | 0.3598 |
| char320 | 0-<590 | 44 | > 45 | 0.105 | 0.1416 | 0.1396 |
| char320 | 590-<2055 | 45 | > 46 | 0.105 | 0.1416 | 0.1396 |
| char320 | 2055-<8405 | 46 | > 47 | -0.088 | -0.0318 | -0.0334 |
| char320 | 8405-<16960 | 47 | > 48 | -0.120 | -0.0756 | -0.0761 |
| char320 | 16960-<20000 | 48 | > 49 | -0.120 | -0.0756 | -0.0761 |
| char320 | 20000-<30000 | 49 | > 50 | -0.213 | -0.2563 | -0.2543 |
| char320 | 30000-<40375 | 50 | > 51 | -0.361 | -0.4934 | -0.4880 |
| char320 | 40375-<70000 | 51 | > 52 | -0.361 | -0.6377 | -0.6277 |
| char320 | 70000-High | 52 | | -0.361 | -0.6579 | -0.6424 |
| char320 | NO INFORMATION | 53 | " = 0 " | 0.000 | 0 | 0.0000 |
| char330 | 0 | 54 | > 55 | 0.251 | 0.2592 | 0.2580 |
| char330 | 1-<250 | 55 | > 56 | -0.144 | -0.1417 | -0.1408 |
| char330 | 250-High | 56 | | -0.343 | -0.3644 | -0.3628 |
| char330 | NO INFORMATION | 57 | " = 0 " | 0.000 | 0 | 0.0000 |
| char380 | -9999999-<0 | 58 | | 0.000 | -0.0000 | 0.0000 |
| char380 | 0-<635 | 59 | > 60 | 0.106 | 0.1128 | 0.1124 |
| char380 | 635-<1210 | 60 | > 61 | -0.014 | -0.0207 | -0.0206 |
| char380 | 1210-<1915 | 61 | > 62 | -0.014 | -0.0207 | -0.0206 |
| char380 | 1915-<5000 | 62 | > 63 | -0.332 | -0.3490 | -0.3474 |
| char380 | 5000-High | 63 | | -0.775 | -0.8131 | -0.8108 |
| char380 | NO INFORMATION | 64 | " = 0 " | 0.000 | 0 | 0.0000 |
| char471 | -9999999 | 65 | | 0.000 | 0.0000 | 0.0000 |
| char471 | 0 | 66 | < 67 | -0.429 | -0.4315 | -0.4297 |
| char471 | 1-<101 | 67 | | 0.016 | 0.0160 | 0.0159 |
| char471 | NO INFORMATION | 68 | " = 0 " | 0.000 | -0.0000 | 0.0000 |
| char503 | 0 | 69 | | 0.010 | 0.0114 | 0.0114 |
| char503 | 1-High | 70 | < 69 | -1.482 | -1.7091 | -1.7003 |
| char503 | NO INFORMATION | 71 | " = 0 & < 69 " | 0.000 | 0 | 0.0000 |
| char533 | -9999999-<1 | 72 | | 0.156 | 0.1661 | 0.1653 |
| char533 | 1 | 73 | > 74 | -0.359 | -0.3725 | -0.3709 |
| char533 | 2 | 74 | > 75 | -0.849 | -0.8236 | -0.8212 |
| char533 | 3 | 75 | >76 | -0.909 | -1.0465 | -1.0400 |
| char533 | 4 | 76 | >77 | -0.909 | -1.0465 | -1.0400 |
| char533 | 5-High | 77 | | -0.909 | -1.0465 | -1.0400 |
| char533 | NO INFORMATION | 78 | " = 0 " | 0.000 | 0 | 0.0000 |
| char635 | 0 | 79 | | 0.004 | -0.0007 | -0.0007 |
| char635 | 1-<3 | 80 | > 81 | 0.050 | 0.0571 | 0.0571 |
| char635 | 3 | 81 | > 82 | 0.050 | 0.0571 | 0.0571 |
| char635 | 4 | 82 | > 83 | -0.153 | -0.0930 | -0.0943 |
| char635 | 5 | 83 | > 84 | -0.400 | -0.3862 | -0.3851 |
| char635 | 6-High | 84 | | -0.619 | -0.5508 | -0.5494 |
| char635 | NO INFORMATION | 85 | " = 0 " | 0.000 | 0 | 0.0000 |
| char665 | 0 | 86 | > 87 | 0.106 | 0.1130 | 0.1125 |
| char665 | 1 | 87 | >88 | -0.529 | -0.5524 | -0.5501 |
| char665 | 2-High | 88 | | -0.572 | -0.6373 | -0.6341 |
| char665 | NO INFORMATION | 89 | " = 0 " | 0.000 | 0 | 0.0000 |
| char710 | -9999999 | 90 | " = 0 " | 0.000 | 0.0000 | 0.0000 |
| char710 | -9999998 | 91 | " = 0 " | 0.000 | 0.0000 | 0.0000 |
| char710 | 0 | 92 | > 93 | 0.066 | 0.0744 | 0.0741 |
| char710 | 1-<360 | 93 | > 94 | 0.066 | 0.0744 | 0.0741 |
| char710 | 360-<675 | 94 | >95 | 0.043 | 0.0327 | 0.0327 |
| char710 | 675-<2435 | 95 | > 96 | -0.343 | -0.3813 | -0.3795 |
| char710 | 2435-High | 96 | | -0.343 | -0.3813 | -0.3795 |
| char710 | NO INFORMATION | 97 | " = 0 " | 0.000 | 0 | 0.0000 |
| char830 | 0 | 98 | > 99 | 0.010 | 0.0114 | 0.0114 |
| char830 | 1 | 99 | > 100 | 0.009 | 0.0114 | 0.0114 |
| char830 | 2-High | 100 | | -0.353 | -0.4092 | -0.4071 |
| char830 | NO INFORMATION | 101 | " = 0 " | 0.000 | 0 | 0.0000 |
| char835 | 0 | 102 | . 103 | 0.056 | 0.0636 | 0.0633 |
| char835 | 1 | 103 | >104 | -0.339 | -0.3758 | -0.3741 |
| char835 | 2 | 104 | > 105 | -0.420 | -0.4760 | -0.4737 |
| char835 | 3 | 105 | > 106 | -0.566 | -0.6313 | -0.6267 |
| char835 | 4-High | 106 | | -0.911 | -1.2386 | -1.2254 |
| char835 | NO INFORMATION | 107 | " = 0 " | 0.000 | 0 | 0.0000 |
| char840 | 0 | 108 | > 109 | 0.235 | 0.2539 | 0.2508 |
| char840 | 1 | 109 | > 110 | -0.313 | -0.2032 | -0.2085 |
| char840 | 2 | 110 | > 111 | -1.062 | -1.1682 | -1.1643 |
| char840 | 3-High | 111 | | -1.497 | -2.3472 | -2.2563 |
| char840 | NO INFORMATION | 112 | " = 0 " | 0.000 | 0 | 0.0000 |
| char843 | 1 | 113 | | 0.737 | 0.8190 | 0.8108 |
| char843 | 2 | 114 | | -0.188 | -0.2589 | -0.2551 |
| char843 | 3 | 115 | | 0.001 | -0.0010 | -0.0009 |
| char843 | 4 | 116 | | -0.013 | 0.0079 | 0.0073 |
| char843 | NO INFORMATION | 117 | " = 0 " | 0.000 | 0 | 0.0000 |
| char860 | 0 | 118 | > 119 | 0.010 | 0.0114 | 0.0114 |
| char860 | 1-High | 119 | | -0.528 | -0.6086 | -0.6055 |
| char860 | NO INFORMATION | 120 | " = 0 " | 0.000 | 0 | 0.0000 |
| char870 | 0 | 121 | > 122 | 0.010 | 0.0114 | 0.0114 |
| char870 | 1 | 122 | > 123 | -0.289 | -0.3336 | -0.3319 |
| char870 | 2-High | 123 | | -0.289 | -0.3336 | -0.3319 |
| char870 | NO INFORMATION | 124 | " = 0 " | 0.000 | 0 | 0.0000 |
| char950 | -9999998-<7011 | 125 | > 126 | 0.659 | 0.6632 | 0.6613 |
| char950 | 3300-<4901 | 126 | > 127 | 0.324 | 0.2570 | 0.2575 |
| char950 | Travel | 127 | > 128 | 0.324 | 0.2570 | 0.2575 |
| char950 | 5511-High | 128 | > 129 | 0.324 | 0.2570 | 0.2575 |
| char950 | MOTO | 129 | > 130 | -0.007 | -0.0542 | -0.0533 |
| char950 | 5697-<7995 | 130 | > 131 | -0.079 | -0.0904 | -0.0907 |
| char950 | 3723-<5945 | 131 | > 132 | -0.079 | -0.0904 | -0.0907 |
| char950 | 5611-<8000 | 132 | > 133 | -0.079 | -0.0904 | -0.0907 |
| char950 | 4814-<4830 | 133 | > 134 | -0.079 | -0.0904 | -0.0907 |
| char950 | 5013-<8100 | 134 | > 135 | -0.157 | -0.0904 | -0.0907 |
| char950 | Gas | 135 | > 136 | -0.157 | -0.0904 | -0.0907 |
| char950 | 5655-<5949 | 136 | > 137 | -0.193 | -0.2126 | -0.2118 |
| char950 | 5300-<5942 | 137 | > 138 | -0.193 | -0.2126 | -0.2118 |
| char950 | 5815-<5963 | 138 | > 139 | -0.490 | -0.2786 | -0.2811 |
| char950 | 5423-<5655 | 139 | | -0.734 | -0.7045 | -0.7014 |
| char950 | NO INFORMATION | 140 | " = 0 " | 0.000 | 0 | 0.0000 |
| char960 | Below -1700 | 141 | < 142 | -0.398 | -0.4029 | -0.4008 |
| char960 | -1700-<-800 | 142 | < 143 | 0.073 | 0.0736 | 0.0732 |
| char960 | -800-<-450 | 143 | < 144 | 0.073 | 0.0736 | 0.0732 |
| char960 | " -450-<High " | 144 | | 0.073 | 0.0736 | 0.0732 |
| char960 | NO INFORMATION | 145 | " = 0 " | 0.000 | 0 | 0.0000 |
| char961 | -9999999 | 146 | | -0.211 | -0.2139 | -0.2127 |
| char961 | -3000-<-1700 | 147 | < 148 | -0.385 | -0.4394 | -0.4371 |
| char961 | -1700-<-800 | 148 | < 149 | -0.048 | 0.0121 | 0.0108 |
| char961 | -800-<550 | 149 | < 150 | 0.148 | 0.1262 | 0.1260 |
| char961 | 550-High | 150 | | 0.186 | 0.1883 | 0.1880 |
| char961 | NO INFORMATION | 151 | " = 0 " | 0.000 | 0 | 0.0000 |
| char962 | Below -1500 | 152 | < 153 | -0.085 | -0.0756 | -0.0758 |
| char962 | -1500-<-1100 | 153 | < 154 | -0.085 | -0.0756 | -0.0758 |
| char962 | -1100-<-850 | 154 | < 155 | -0.073 | -0.0756 | -0.0758 |
| char962 | -850-<-550 | 155 | < 156 | -0.073 | -0.0756 | -0.0758 |
| char962 | -550-<-400 | 156 | < 157 | 0.019 | 0.0170 | 0.0173 |
| char962 | -400-<-300 | 157 | < 158 | 0.019 | 0.0170 | 0.0173 |
| char962 | -300-<1 | 158 | < 159 | 0.019 | 0.0170 | 0.0173 |
| char962 | 1-<200 | 159 | < 160 | 0.104 | 0.0967 | 0.0969 |
| char962 | 200-High | 160 | | 0.174 | 0.1740 | 0.1739 |
| char962 | NO INFORMATION | 161 | " = 0 " | 0.000 | 0 | 0.0000 |
| char965 | Below -950 | 162 | < 163 | -0.328 | -0.2923 | -0.2912 |
| char965 | -950-<-750 | 163 | < 164 | -0.328 | -0.2923 | -0.2912 |
| char965 | -750-<-550 | 164 | < 165 | -0.321 | -0.2923 | -0.2912 |
| char965 | -550-<-400 | 165 | < 166 | -0.321 | -0.2923 | -0.2912 |
| char965 | -400-<-300 | 166 | < 167 | -0.044 | -0.0313 | -0.0315 |
| char965 | -300-<-200 | 167 | < 168 | 0.186 | 0.1809 | 0.1797 |
| char965 | -200-<-100 | 168 | < 169 | 0.201 | 0.1809 | 0.1797 |
| char965 | -100-<80 | 169 | < 170 | 0.366 | 0.3245 | 0.3236 |
| char965 | 80-High | 170 | | 0.366 | 0.3245 | 0.3236 |
| char965 | NO INFORMATION | 171 | " = 0 " | 0.000 | 0 | 0.0000 |